\documentclass[twocolumn,preprintnumbers,amsmath,amssymb]{revtex4}

\usepackage{graphicx}
\usepackage{dcolumn}
\usepackage{bm}

\begin{document}


\title{Magnetic phase transitions in NdCoAsO}

\author{Michael A. McGuire}
\affiliation{Oak Ridge National Laboratory, Oak Ridge, Tennessee 37831 USA}
\author{Delphine J. Gout}
\affiliation{Oak Ridge National Laboratory, Oak Ridge, Tennessee 37831 USA}
\author{V. Ovidiu Garlea}
\affiliation{Oak Ridge National Laboratory, Oak Ridge, Tennessee 37831 USA}
\author{Athena S. Sefat}
\affiliation{Oak Ridge National Laboratory, Oak Ridge, Tennessee 37831 USA}
\author{Brian C. Sales}
\affiliation{Oak Ridge National Laboratory, Oak Ridge, Tennessee 37831 USA}
\author{David Mandrus}
\affiliation{Oak Ridge National Laboratory, Oak Ridge, Tennessee 37831 USA}

\date{\today}

\begin{abstract}
NdCoAsO undergoes three magnetic phase transitions below room temperature. Here we report the results of our experimental investigation of this compound, including determination of the crystal and magnetic structures using powder neutron diffraction, as well as measurements of electrical resistivity, thermal conductivity, Seebeck coefficient, magnetization, and heat capacity. These results show that upon cooling a ferromagnetic state emerges near 69 K with a small saturation moment of $\sim 0.2 \mu_B$, likely on Co atoms. At 14 K the material enters an antiferromagnetic state with propagation vector (0 0 1/2) and small ordered moments ($\sim 0.4 \mu_B$) on Co and Nd. Near 3.5 K a third transition is observed, and corresponds to the antiferromagnetic ordering of larger moments on Nd, with the same propagation vector. The ordered moment on Nd reaches 1.39(5)$\mu_B$ at 300 mK. Anomalies in the magnetization, electrical resistivity, and heat capacity are observed at all three magnetic phase transitions.
\end{abstract}

\maketitle

\section{Introduction}

Iron based superconductors have been discovered in several families of compounds containing square nets of iron with formal valence near 2+ in tetrahedral coordination. These include the 1111 family based on LaFeAsO \cite{Kamihara2008}, the 122 family based on BaFe$_2$As$_2$ \cite{BaFe2As2-SC}, the 111 family based on LiFeAs \cite{LiFeAs}, the 11 family based on Fe$_{1+\delta}$Se \cite{FeSe-SC}, and the 32522 family based on Sr$_3$Sc$_2$O$_5$Fe$_2$As$_2$ \cite{32522-SC}. In the 1111 and 122 systems substitution of cobalt for iron has been demonstrated as an effective way to induce superconductivity \cite{Sefat-Co1111, Sefat-Co122}. While the main interest in these materials is the superconducting state, non-superconducting members of these families have also shown interesting behavior, and their study may help further our understanding of the superconducting materials. Upon cooling, the pure iron compounds LaFeAsO and BaFe$_2$As$_2$ undergo a spin density wave transition with an accompanying crystallographic distortion \cite{BaFe2As2, McGuire-LaFeAsO, Nature-LaFeAsO}, a ground state that seems to compete with superconductivity. In the pure cobalt compound LaCoAsO, small moments on Co order ferromagnetically below about 60 K with saturation moments of 0.3$-$0.5 $\mu$B per Co \cite{Yanagi-LaCoAsO, Sefat-Co1111, Ohta-LaCoAsO}. Experimental and theoretical studies suggest that BaCo$_2$As$_2$ is a highly renormalized paramagnetic metal, near a ferromagnetic instability \cite {Sefat-BaCo2As2}. It has been proposed that spin fluctuations play an important role in the magnetic behavior of LaCoAsO \cite{Yanagi-LaCoAsO, Ohta-LaCoAsO}, as well as the magnetic and superconducting properties of the iron-based superconductors \cite{Ning}.

In the superconducting 1111 materials, the transition temperature T$_c$ is increased significantly as La is replaced by other lanthanides, from 28 K for LaFeAsO$_{1-x}$F$_x$ to 55 K for SmFeAsO$_{1-x}$F$_x$ \cite{Sefat-F1111, SmFeAsOF}. It has also been shown that the identity of the lanthanide $Ln$ strongly effects the low temperature transport properties of the undoped parent compounds $Ln$FeAsO, and interaction between the $Ln$ and Fe magnetism has been observed below the $Ln$ antiferromagnetic ordering temperatures \cite{McGuire-LnFeAsO}. To further probe the interesting magnetic properties that are found in these layered itinerant magnetic materials, we have begun investigating the effects of replacing La in LaCoAsO with magnetic lanthanides.

Here we report our results for NdCoAsO. It is isostructural to LaFeAsO and LaCoAsO, adopting the tetragonal ZrCuSiAs structure type with space group \textit{P4/nmm }\cite{Quebe}. Our magnetization measurements revealed upon cooling a series of magnetic transitions: a ferromagnetic transition similar to that reported for LaCoAsO, and two antiferromagnetic transitions at lower temperatures. During the course of this work, a report of magnetization measurements on NdCoAsO has been published \cite{Ohta-LCoAsO}. Those results are in good agreement with the results presented here; however, the authors discussed only two of the three magnetic phase transitions. In addition to magnetization analysis, we have used powder neutron diffraction to determine the antiferromagnetic structure of the Nd and Co moments in this material. We also report detailed crystal structure information as well as the effects of the phase transitions on the transport properties and heat capacity.

\section{Experimental Details}

Polycrystalline samples of NdCoAsO were prepared by solid state reactions, using procedures similar to those reported for this and other related compounds \cite{Sefat-Co1111, McGuire-LaFeAsO, McGuire-LnFeAsO, Ohta-LCoAsO, Yanagi-LaCoAsO}. CoAs was first prepared from Co powder and As pieces heated very slowly to 700$^\circ$C and then 1065$^\circ$C. Nd$_2$O$_3$, Nd, and CoAs powders were then thoroughly mixed in an agate mortar and pestle inside a helium filled glovebox. The mixtures were pressed into pellets and sealed in silica tubes which were evacuated and backfilled with about 0.2 atm argon gas. The samples were reacted at 1200$^\circ$C for 12 hours, reground and re-pelletized, and heated for a second time at 1200$^\circ$C for 12 hours. Starting materials were of 99.9\% purity or better. From powder X-ray and neutron diffraction analysis we estimate the purity of our NdCoAsO samples to be $\gtrsim$ 95\%. Observed impurity phases were CoAs (an antiferromagnetic with T$_N$ = 54 K \cite{Lewis-CoAs}) and Nd$_2$SiO$_5$. No contributions from these impurity phases to the measured physical properties were observed.

Heat capacity and transport measurements were performed using a Quantum Design Physical Property Measurement System. SQUID magnetometery measurements were carried out using a Quantum Design Magnetic Property Measurement System. Neutron powder diffraction data were collected from a 9 gram sample of NdCoAsO at temperatures from 300 mK to 300 K on the High-Resolution Powder Diffractometer (HB2A) at the High Flux Isotope Reactor at Oak Ridge National Laboratory, using 12'-31'-6' collimation and with a wavelength of 1.538 {\AA} Ge(115). More details about the HB2A instrument and data collection strategies can be found in Ref.~\onlinecite{HB2A}. The crystallographic structure of NdCoAsO was refined at 300 K, 120 K and 30 K (above the antiferromagnetic ordering temperatures) using the JANA2006 program package \cite{JANA, JANA2}. Effects of the ferromagnetic ordering expected at 30 K could not be reliably observed in the current data, due to the small magnitudes of the moments and their coincidence with the much stronger nuclear Bragg reflections. JANA2006 was also used for the determination of the magnetic structure at lower temperatures (20 K and below). Refinement of the magnetic structure was also performed using the program Fullprof \cite{Fullprof}. These two software packages use different approaches to describe the magnetic structure, and the two programs gave equivalent results. The values of magnetic moments and magnetic refinement results presented in this report were obtained using Fullprof.

\section{Results and Discussion}

\subsection{Magnetization}

\begin{figure}
\includegraphics[width=3.0in]{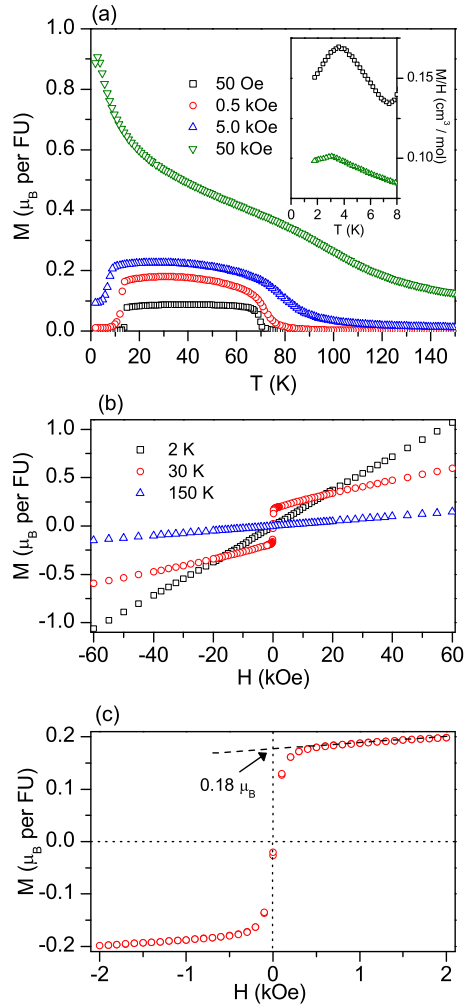}
\caption{\label{fig:mag}
(color online) Magnetic properties of NdCoAsO. (a) Temperature dependence of the magnetization of NdCoAsO at four different applied magnetic fields. (b) Magnetic field dependence of the magnetization at three different temperatures, showing a component which saturates at 30 K, near the center of the ferromagnetic-like temperature regime, and linear behavior in both the paramagnetic state at higher temperature (150 K) and the antiferromagnetic state at lower temperature (2 K). (c) Low field behavior at 30 K which indicates a saturation moment near 0.18 $\mu_B$ per formula unit (FU).
}
\end{figure}

Figure \ref{fig:mag} shows the measured magnetic properties of NdCoAsO. These results are consistent with the report of Ohta and Yoshimura \cite{Ohta-LCoAsO}. Three phase transitions are observed in the temperature dependence of the magnetization (Figure \ref{fig:mag}a). Upon cooling in low applied fields, an abrupt increase is observed near T$_C$ = 69 K, suggestive of ferromagnetic ordering. This is followed by an antiferromagnetic transition at T$_{N1}$ = 14 K, at which point the magnetization sharply decreases (except in the strongest applied fields). A third anomaly, a downturn in the magnetization upon cooling, is observed near T$_{N2}$ = 3.5 K (Figure \ref{fig:mag}a, inset). At high fields (50 kOe) much of this behavior is overwhelmed by Nd paramagnetism; however, a broad feature can be discerned below 100 K, and a sharp cusp is observed at 3 K. The ferromagnetic-like region is extended to cover a broader temperature range as the applied field is increased. A Curie-Weiss fit to data collected at 50 kOe over the temperature range of 200$-$300 K (not shown) gives an effective moment of 3.5 $\mu_B$, close to the expected value of 3.62 $\mu_B$ for Nd$^{3+}$, and in good agreement with the literature report for this material \cite{Ohta-LCoAsO}. If the magnetism in the Co layer is  itinerant in nature, a Co contribution to this effective moment may not be expected. However, Curie-Weiss behavior has been observed in LaCoAsO \cite{Ohta-LaCoAsO, Sefat-Co1111} with an effective moment near 1.3 $\mu_B$ per Co. Thus, the effective moment observed here for NdCoAsO may include contributions from both magnetic ions. The fitted Weiss temperature for NdCoAsO is 34 K. Interestingly, this value is positive, which would indicate predominantly ferromagnetic interaction at high temperatures.

The field dependence of the magnetization at 2, 30, and 150 K is shown in Figure \ref{fig:mag}b. Linear behavior is observed at 2 and 150 K. At 30 K, near the middle of the ferromagnetic region, a rapid saturation is observed up to 0.5 kOe, followed by linear region up to 60 kOe. No magnetic hysteresis is observed. A linear fit to the data above 1 kOe reveals a saturated moment of 0.18 $\mu_B$ per formula unit (Figure \ref{fig:mag}c). A similar saturation moment was observed by Ohta and Yoshimura \cite{Ohta-LCoAsO}. Since similar behavior has been observed in LaCoAsO, where Co is the only magnetic atom, it is likely that the saturated component in NdCoAsO arises from Co, and the linear behavior at higher fields originates primarily from the paramagnetic response of Nd. The sharp decrease in magnetization at T$_{N1}$ likely indicates a ferromagnetic-antiferromagnetic transition, as noted previously \cite{Ohta-LCoAsO}. The downturn in magnetization near 3 K suggests antiferromagnetic order of Nd moments. This occurs near 2 K in the closely related compound NdFeAsO \cite{NdFeAsO-Ndmag}. The evolution of the magnetic structure of this material upon cooling through these magnetic phase transitions, determined from neutron diffraction measurements, is discussed below.

\subsection{Neutron diffraction}

\begin{table}
  \caption{Crystallographic parameters and agreement factors (R$_{obs}$ and goodness of fit) from Rietveld refinements in space group \textit{P4/nmm} with Nd atoms at (1/4, 1/4, z-Nd), Co atoms at (1/4, 3/4, 1/2), As atoms at (1/4, 1/4, z-As), and O atoms at (1/4, 3/4, 0). U$_{iso}$ are isotropic displacement parameters.}
\begin{tabular*}{3.0in}%
    {@{\extracolsep{\fill}}lccc}
  \hline
  T (K) & 300 & 120 & 30 \\
  &&&\\
  a ({\AA}) & 3.98423(8) & 3.98299(8) & 3.98237(7) \\
  c ({\AA}) & 8.3333(3) & 8.3193(2) & 8.3123(2) \\
  &&&\\
  z-Nd & 0.1422(2) & 0.1421(4) & 0.1418(2) \\
  z-As & 0.6501(3) & 0.6507(3) & 0.6511(3) \\
  &&&\\
  U$_{iso}$-Nd ({\AA}$^2$) & 0.0123(5) & 0.0041(5) & 0.0019(5) \\
  U$_{iso}$-Co ({\AA}$^2$) & 0.0187(11) & 0.0058(12) & 0.0036(11) \\
  U$_{iso}$-As ({\AA}$^2$) & 0.0198(6) & 0.0080(6) & 0.0061(5) \\
  U$_{iso}$-O  ({\AA}$^2$) & 0.017(6) & 0.0088(6) & 0.0079(6) \\
  &&&\\
  R$_{obs} (\%)$ & 1.97 & 1.72 & 1.68 \\
  GoF & 1.73 & 1.47 & 1.98 \\
  \hline
\end{tabular*}
\label{table:structure}
\end{table}
\begin{figure}
\includegraphics[width=3.5in]{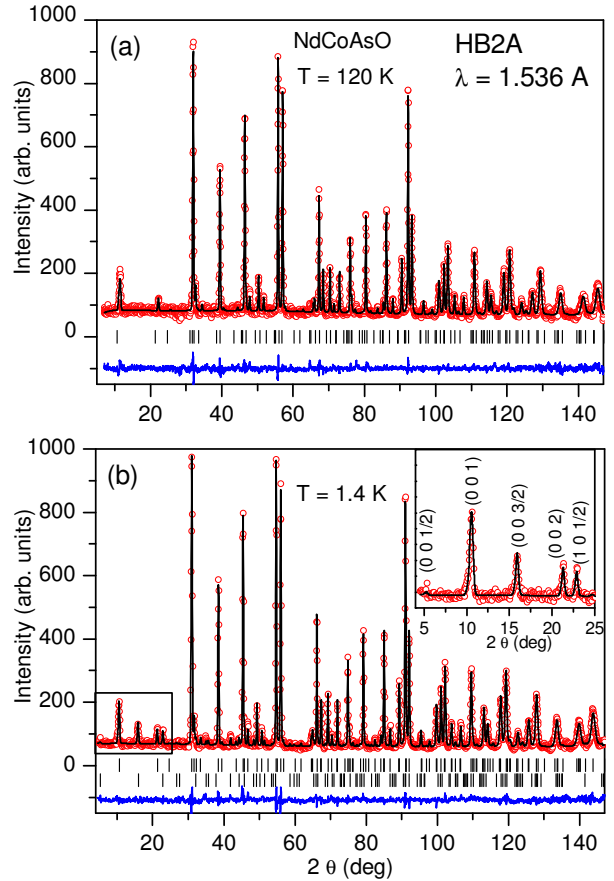}
\caption{\label{fig:Rietveld}
(color online) Rietveld refinements of powder neutron diffraction data at (a) T = 120 K and (b) T = 1.4 K showing measured data (circles) and fitted and difference curves. Tick marks locate Bragg reflections. In (b) upper ticks represent nuclear reflections and lower ticks indicate magnetic reflections. The inset in (b) shows the low angle region of the 1.4 K data, with peaks labeled by their indices.}
\end{figure}

Results of crystal structure refinements at three temperatures above the antiferromagnetic transitions are reported in Table \ref{table:structure}. The diffraction pattern at T = 120 K is displayed in Figure \ref{fig:Rietveld}a. The tetragonal space group \textit{P4/nmm} was used at all temperatures. We note that the nuclear Bragg reflections at lower temperatures were also well described by this space group (Figure \ref{fig:Rietveld}b). The variation of Bragg peak widths with temperature gave no indication of a structural distortion within the resolution of the current data. As Table \ref{table:structure} indicates, no unusual behavior in the lattice parameters, atomic positions, or isotropic displacement parameter are observed upon cooling from room temperature into the ferromagnetic state.

\begin{figure}
\includegraphics[width=3.0in]{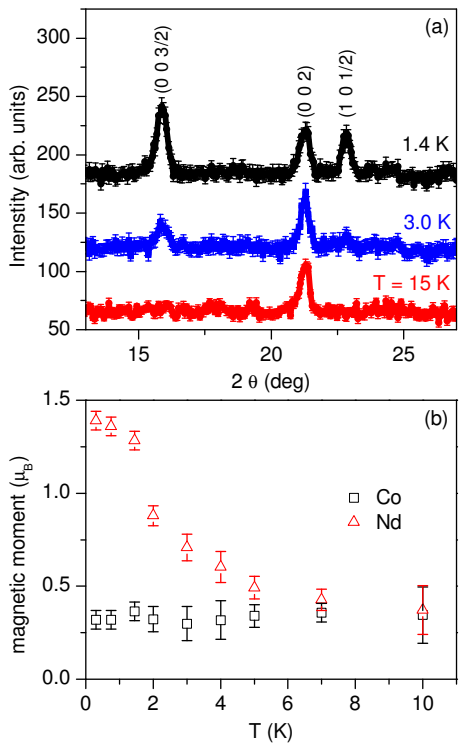}
\caption{\label{fig:NPD-mag}
(color online) (a) Low angle neutron powder diffraction data from NdCoAsO at 15 K, 3 K, and 1.4 K. The (0 0 2) nuclear Bragg peak appears at 21.3 degrees. Magnetic reflections occur at 15.9 and 22.8 degrees. Data collected at different temperatures are offset vertically for clarity. (b) The temperature dependence of the refined antiferromagnetically ordered moments on Co and Nd for T $\leq$ 10 K.
}
\end{figure}
\begin{figure}
\includegraphics[width=2.5in]{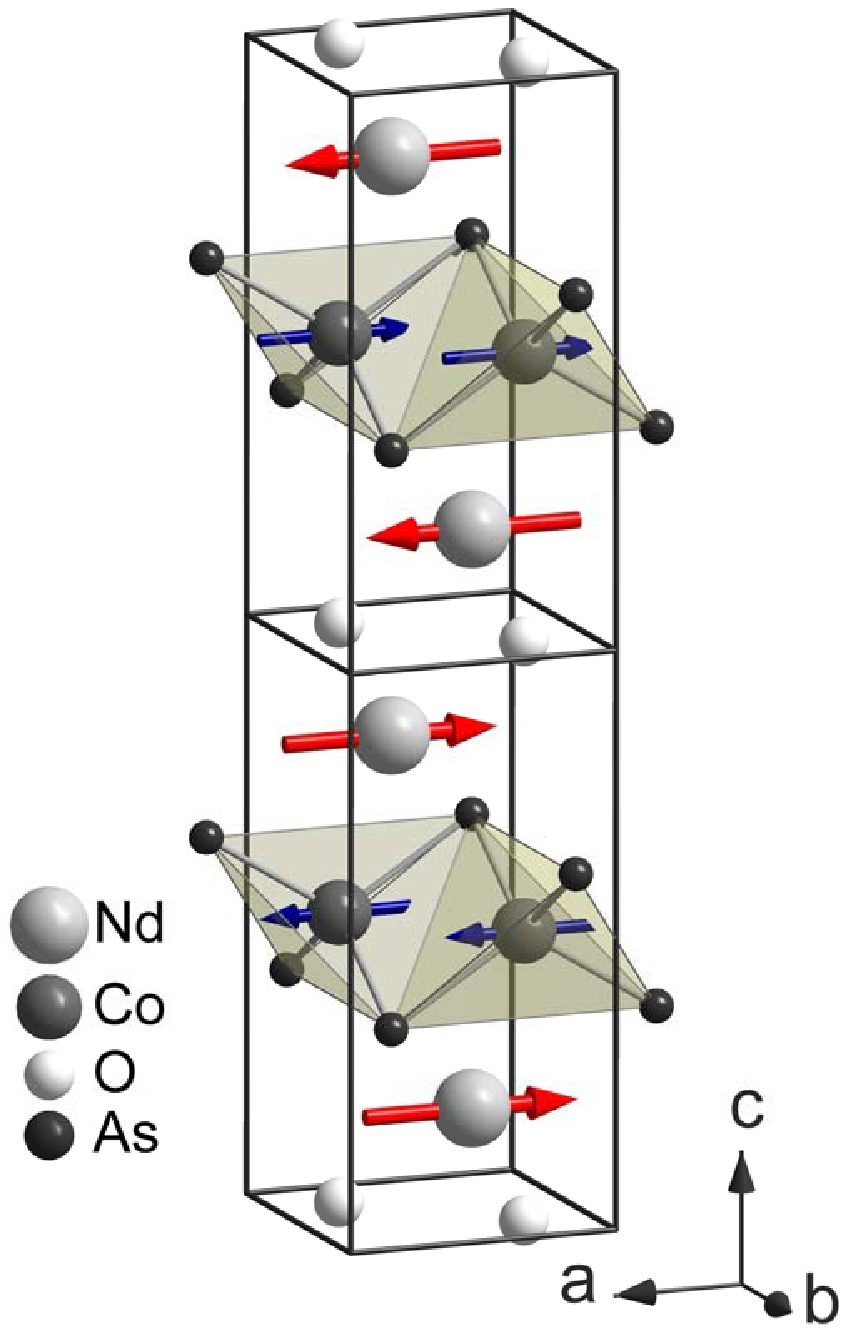}
\caption{\label{fig:structure}
(color online) Crystal and magnetic structure of NdCoAsO in the antiferromagnetic states. Arrows indicate magnetic moments on Nd and Co which lie along the a-axis.
}
\end{figure}

Low angle neutron diffraction patterns collected at 15 K, 3 K, and 1.4 K are shown in Figure \ref{fig:NPD-mag}a. In the angular range displayed, only the nuclear (002) peak is observed at 15 K. At lower temperatures the (0 0 3/2) and (1 0 1/2) magnetic reflections are observed. Representational analysis~\cite{Bertaut1, Bertaut2, Bertaut3, Bertaut4} has been used to determine the symmetry-allowed magnetic structures, given the crystal structure and the propagation vector of the magnetic ordering. The calculations were carried out using the program SARA{\textit{h}}-Representational Analysis.\cite{Sarah} They involve the determination of the space group symmetry elements, $g$, that leave the propagation vector ${\mathbf k}$ invariant, forming the little group $G_{\mathbf k}$.~\cite{Kovalev} The decomposition of the magnetic representation in terms of the non-zero irreducible representations (IRs) of $G_{\textbf{k}}$ for each crystallographic site examined (Co and Nd), and their associated basis vectors, are given in Table \ref{basis-vectors}.

\begin{table}[h]
\caption{Basis vectors for the space group P 4/n m m:2 with ${\bf k}_{19}=( 0,~0,~.5)$. The Co atoms of the nonprimitive basis are defined according to 1: $( .75,~ .25,~ .5)$, 2: $( .25,~ .75,~ .5)$, while the Nd atoms are
1: $( .25,~ .25,~ .141)$, 2: $( .75,~ .75,~ .858)$.}
\label{basis-vectors}
\begin{tabular}{|cc|ccc|cc|ccc|}
   \multicolumn{5}{c}{Co sites}& \multicolumn{5}{c}{Nd sites}\\
   IR   & Atom      & $m_{\|a}$ & $m_{\|b}$ & $m_{\|c}$ & IR  & Atom   & $m_{\|a}$ & $m_{\|b}$ & $m_{\|c}$ \\
\hline
$\Gamma_{2}$ &      1 &      0 &      0 &      1 &  $\Gamma_{2}$ &      1 &    0 &      0 &      1  \\
             &      2 &      0 &      0 &      1 &               &      2 &    0 &      0 &      1  \\
$\Gamma_{7}$ &      1 &      0 &      0 &      1 &  $\Gamma_{3}$ &      1 &    0 &      0 &      1  \\
             &      2 &      0 &      0 &     -1 &               &      2 &    0 &      0 &     -1  \\
$\Gamma_{9}$ &      1 &      1 &      0 &      0 &  $\Gamma_{9}$ &      1 &    1 &      0 &      0  \\
             &      2 &     -1 &      0 &      0 &               &      2 &   -1 &      0 &      0  \\
             &      1 &      0 &      1 &      0 &               &      1 &    0 &     -1 &      0  \\
             &      2 &      0 &     -1 &      0 &               &      2 &    0 &      1 &      0  \\
$\Gamma_{10}$ &      1 &      0 &      1 &      0 &  $\Gamma_{10}$ &     1 &   0 &      1 &      0  \\
             &      2 &      0 &      1 &      0 &                &      2 &   0 &      1 &      0  \\
             &      1 &      1 &      0 &      0 &                &      1 &   1 &      0 &      0  \\
             &      2 &      1 &      0 &      0 &                &      2 &   1 &      0 &      0  \\
\end{tabular}
\end{table}

Since the neutrons sense only the projections of the magnetic moments in the plane perpendicular to the scattering vector, the presence of a strong (0 0 3/2) reflection is a clear indication that the moments exhibit components lying in the basal plane. Thus, the basis vectors involved in the magnetic structure appeared to be limited to those associated with the IRs $\Gamma_{9}$ and $\Gamma_{10}$. Several magnetic structure models were tested. It was found that the simplest model which would fit the data adequately is one consisting of Co magnetic moments parallel to the tetragonal $a$-axis, which are compensated by Nd moments pointing in opposite direction. Such a model corresponds to the IR $\Gamma_{10}$, that allows ferromagnetic alignment within each magnetic sublattices (Nd or Co), as well as an antiferromagnetic coupling between them. Such a moment configuration corresponds to the magnetic space group $Pcc'n$. A view of the magnetic structure is displayed in Figure \ref{fig:structure}. The Co magnetic moments alternate their directions from layer to layer, so that the magnetic unit cell is twice as large as the chemical cell. It is important to note that the (0 0 1/2) magnetic reflection is much weaker than (0 0 3/2) at all temperatures investigated (see Figure \ref{fig:Rietveld}b). This strongly supports the presence of magnetism on both magnetic ions. Analysis of the 300 mK data yielded the ordered moments: $m_{Co}\approx0.32(5)~\mu _{B}$ and $m _{Nd}\approx1.39(5)~\mu _{B}$.

The temperature dependences of the magnitude of the ordered moments on Co and Nd at T $\leq$ 10 K determined from neutron powder diffraction refinements are shown in Figure \ref{fig:NPD-mag}b. Non-zero values of these moments are not detected at 15 K and above. The magnitude of the ordered moment on Co is temperature independent below T$_{N1}$.  The magnitude of the ordered moment on Nd increases slowly with decreasing temperature for $T_{N2} < T < T_{N1}$. Below $T_{N2}$, the Nd ordered moment abruptly increases, and is nearly saturated at the lowest temperature investigated here (300 mK). These results suggest that $T_{N1}$ involves an antiferromagnetic ordering of moments on cobalt, likely a reorientation of the moments which are ordered ferromagnetically at higher temperatures, which induces a partial ordering of Nd magnetic moments. Antiferromagnetic ordering of full Nd moments then sets in below $T_{N2}$. This behavior is quite similar to that observed in Nd$_2$CuO$_4$, in which moments on Nd are induced below one of the Cu spin reorientation temperatures (30 K) and then order at low temperatures ($\sim$ 1 K) \cite{Nd2CuO4}.

\subsection{Heat capacity}

\begin{figure}
\includegraphics[width=3.0in]{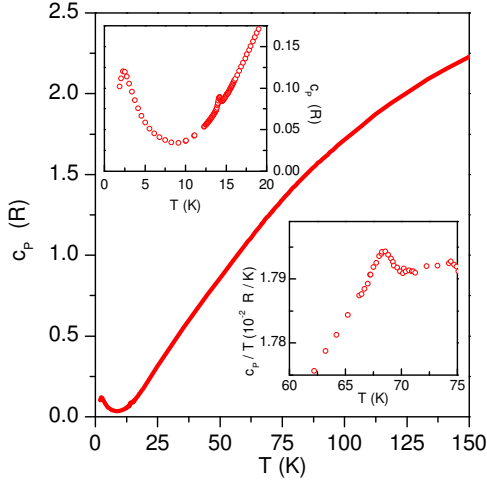}
\caption{\label{fig:cP}
(color online) (a) Specific heat c$_P$ of NdCoAsO in units of R per mole of atoms. The insets show the behavior of c$_P$ near the phase transitions. The anomaly near 69 K is emphasized by plotting c$_{P}$/T vs. T.
}
\end{figure}

Results of heat capacity measurements are presented in Figure \ref{fig:cP}. The temperature dependence of the specific heat $c_P$ per mole of atoms is shown in the main panel for temperatures below 150 K. At room temperature, $c_P$ reaches a value of 2.9 R. Heat capacity anomalies associated with magnetic phase transitions are emphasized in the insets of Figure \ref{fig:cP}. The feature at 69 K is subtle, and is most easily observed in a plot of $c_P/T$ vs. T. Another small peak is observed at 13 K, and a large anomaly occurs near 3.5 K.

The peaks in specific heat shown in Figure \ref{fig:cP} were integrated to obtain entropy changes associated with each transition. In the absence of an appropriate non-magnetic analogue for determination of the background lattice/electronic contribution to the heat capacity, analysis of the anomalies was performed using polynomial fits above and below the phase transitions to estimate the non-magnetic portion of the heat capacity. For the peaks at 3.5 K and 14 K, the background was estimated from a single fit to the data from 12.8 to 18 K, excluding the range around the 14 K peak, using the equation $a_1T+a_2T^3$, with fitting coefficients $a_1$ and $a_2$. For the 69 K peak, the background was estimated using a third order polynomial from 60 to 80 K, excluding the temperature range near the peak. The background was subtracted from the data to obtain $\Delta c_P$, and $\Delta c_P/T$ was integrated to obtain the entropy changes $\Delta$S. The integrations result in $\Delta$S = $\sim 1\times10^{-4}\ R$ and $\sim 3\times10^{-4}\ R$ for the transitions at T$_C$ and T$_{N1}$, respectively. The relatively small entropy change associated with the transitions at T$_{C}$ and T$_{N1}$ are consistent with the small sizes of the moments which order at these temperatures.

Since the peak at T$_{N2}$ extends beyond the lowest temperature measured in this work, we can only roughly estimate the entropy associated with this transition. To do this, the low T behavior of the heat capacity anomaly $\Delta c_P$ was interpolated from 0 to 1.87 K. Two interpolation schemes were used: $\Delta c_P = b_1 T$ and $\Delta c_P = b_2 T^3$. The integrated entropy change is of course sensitive to the interpolation scheme, giving $\Delta$S = 0.21 $R$ for linear interpolation and 0.14 $R$ for cubic interpolation.

\subsection{Transport properties}

\begin{figure}
\includegraphics[width=3.0in]{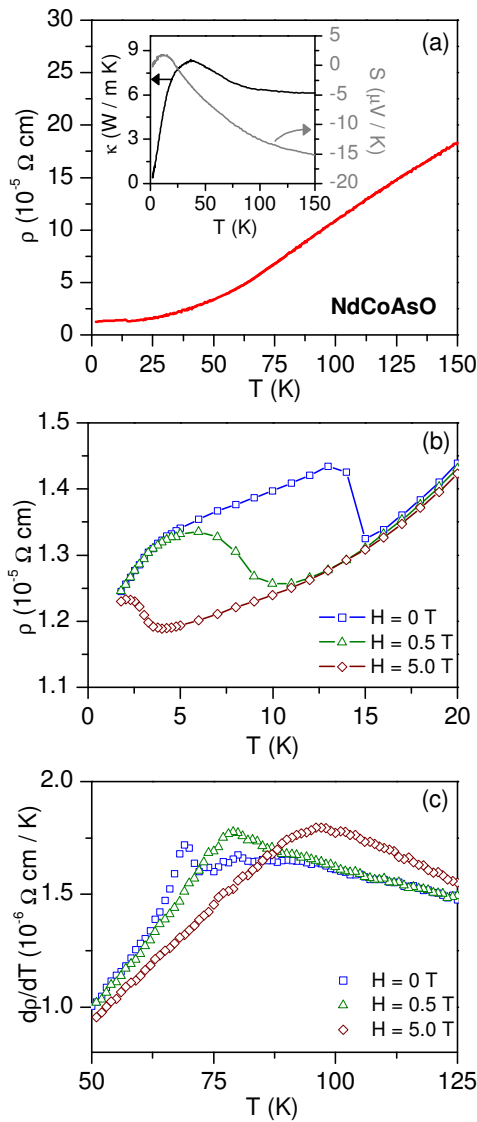}
\caption{\label{fig:transport}
(color online) Transport properties of NdCoAsO. (a) Electrical resistivity $\rho$. The inset shows the Seebeck coefficient and thermal conductivity. (b) Resistivity behavior near the antiferromagnetic phase transitions, including the effect of applied magnetic field. (c) Temperature derivative of the resistivity near the ferromagnetic phase transition and the effects of applied magnetic field.
}
\end{figure}

The temperature dependence from 2 to 150 K of the electrical resistivity $\rho$, Seebeck coefficient $S$, and total thermal conductivity $\kappa$ of a polycrystalline sample of NdCoAsO is shown in Figure \ref{fig:transport}a. The electrical resistivity displays metallic behavior up to room temperature, and reaches a value of 0.36 m$\Omega$ cm at 300 K. The residual resistivity ratio $R(300 K)/R(1.8 K)$ is 29, relatively high for a polycrystalline material. Indications of the low temperature magnetic phase transitions observed in $\rho$ are shown in Figure \ref{fig:transport}b. Upon cooling through T$_{N1}$, a sharp increase in $\rho$ occurs. Since this phase transition involves a reorientation of moments on Co atoms in the conducting CoAs layer, the observed change in $\rho$ is likely due to changes in the magnetic scattering rate of the charge carriers. Below about 4 K a gradual decrease in $\rho$ is seen, and is attributed to the magnetic ordering of larger moments on Nd which occurs near 3 K. A similar decrease in resistivity is also seen in \textit{Ln}FeAsO below the magnetic ordering temperature of the rare earth ions \textit{Ln} \cite{McGuire-LnFeAsO}. The effects of the ferromagnetic transition on $\rho$ are less dramatic, and are best observed in the temperature derivative d$\rho$/dT shown in Figure \ref{fig:transport}c. A local maximum in d$\rho$/dT occurs near T$_C$.

The dependence of the resistivity anomalies on applied magnetic field is also shown in Figure \ref{fig:transport}b and c. An applied field extends the ferromagnetic region to a larger temperature range by pushing T$_{C}$ up and T$_{N1}$ down. This is consistent with the magnetization measurements presented above.

The thermal conductivity (Figure \ref{fig:transport}a, inset) shows behavior typical of crystalline materials. No effects of the phase transitions on $\kappa$ are observed. The Seebeck coefficient (Figure \ref{fig:transport}a, inset) is negative for temperatures above 23 K, indicating predominantly n-type conduction. No effects of the magnetic transitions at T$_{C}$ and T$_{N2}$ can be resolved in the measured behavior of $S$. The maximum that occurs in $S$ near 13 K may be related to the magnetic transition at T$_{N1}$, or may simply be a result of the contribution of multiple bands, including at least one hole band, to the conduction in NdCoAsO.

\section{Conclusions}
We have shown that NdCoAsO undergoes three magnetic phase transitions at low temperatures, and determined the magnetic structure in the antiferromagnetically ordered states. A bulk ferromagnetic phase transition is observed at $T_{C}$ = 69 K by magnetization measurements and heat capacity ($c_P$) measurements, which show a small anomaly at this temperature. Two antiferromagnetic phase transitions are observed, one at $T_{N1}$ = 14 K, with a small $c_P$ anomaly, and another at $T_{N2}$ = 3.5 K, with a larger peak in $c_P$. The observed behavior is attributed to ferromagnetic ordering of small moments on Co at $T_{C}$, a reorientation of these moments into an antiferromagnetically ordered state at $T_{N1}$ with similarly small ordered moments induced on Nd, and a subsequent antiferromagnetic ordering of ``full'' Nd moments at $T_{N2}$. In the antiferromagnetic states, all ordered moments lie in the ab-plane. The Co atoms in each CoAs layer are ferromagnetically ordered, and these layers are ordered antiferromagnetically along the c-direction. The two Nd sites in each NdO layer are arranged antiferromagnetically, and alternate in direction between layers. These arrangements result in a doubling of the chemical unit cell along c, and a propagation vector of (0 0 1/2). No indication of a structural phase transition is observed down to 1.4 K. Seebeck coefficient measurements suggest electron dominated conduction near room temperature, but show small positive values at low temperatures, likely due to multiband effects. Analysis of electrical resistivity reveals anomalies at all three magnetic phase transitions, and show that the ferromagnetic temperature regime is increased in applied magnetic fields, as $T_{C}$ is pushed to high temperature and $T_{N1}$ is lowered, indicating an increased stabilization of the ferromagnetic state.
\\

\textit{Note}: While this manuscript was under review, the authors became aware of another report on NdCoAsO (arXiv:1001.2713v1) which is in general agreement with the results presented here.
\\

We gratefully acknowledge helpful discussions with K. V. Vemuru and V. Petricek. Research supported by the U.S. Department of Energy, Office of Basic Energy Sciences, Division of Materials Sciences and Engineering under Award \# (synthesis and physical property measurements). The work at Oak Ridge National Laboratory's High Flux Isotope Reactor (neutron diffraction) was sponsored by the U.S. Department of Energy, Office of Basic Energy Sciences, Scientific User Facilities Division under Award \# . Part of this research performed by Eugene P. Wigner Fellows at ORNL.

\end{document}